\documentclass[11pt]{article}

\usepackage[draft]{fixme}
\usepackage{acl2010}
\usepackage{times}
\usepackage{url}
\usepackage{latexsym}

\usepackage{amsmath, amssymb, amsthm, xspace, enumerate,nopageno}

\usepackage{tikz}
\usetikzlibrary{automata}

\usepackage[ruled,vlined]{algorithm2e}
\usepackage{mathtools}

\usepackage{relsize}

\allowdisplaybreaks

\newif\iffullversion\fullversionfalse

\newcommand{\pr}[2]{{\it pre}_{#1}(#2)}
\newcommand{\su}[2]{{\it suc}_{#1}(#2)}
\newcommand{\guard}{$\exists r,u,v,w:\label{alg:line:loop-begin}(u,v) \in
\interp{r},w\in S(u),$\\\qquad\quad$\su{r}{w}\cap S(v)=\emptyset$}
\newcommand{\unary}{P}

\newcommand{\FOL}{\ensuremath{\mathcal{FL}}\xspace}
\newcommand{\EPFOL}{\ensuremath{{\FOL^-}}\xspace}
\newcommand{\ALC}{\ensuremath{\mathcal{ALC}}\xspace}
\newcommand{\EL}{\ensuremath{\mathcal{EL}}\xspace}
\newcommand{\ELAN}{\ensuremath{{\mathcal{EL}^+}}\xspace}

\newcommand{\st}{\mathit{\tau}\xspace}

\newcommand\+[1]{\mathcal{#1}}

\newcommand{\simil}[1]{\mathrel{\stackrel{\mathsmaller{#1}}{\raisebox{-.07em}{$\rightsquigarrow$}}}}
\newcommand{\simul}[1]{\mathrel{\stackrel{\mathsmaller{#1}}{\underline{\makebox[.75em][l]{\hspace{-.05em}\raisebox{-.15ex}{$\rightarrow$}}}}}}

\newcommand{\size}[1]{\# #1}

\newcommand{\atomL}{\textsc{atom}_L\xspace}
\newcommand{\atomR}{\textsc{atom}_R\xspace}
\newcommand{\atomLR}{\textsc{atom}_{L/R}\xspace}

\newcommand{\zig}{\textsc{rel}_L\xspace}
\newcommand{\zag}{\textsc{rel}_R\xspace}
\newcommand{\zigzag}{\textsc{rel}_{L/R}\xspace}

\newcommand{\injL}{\textsc{inj}_L\space}
\newcommand{\injR}{\textsc{inj}_R\space}
\newcommand{\injLR}{\textsc{inj}_{L/R}\space}

\newcommand{\diam}{\exists r.}

\newcommand{\pos}{\EL}
\newcommand{\posre}{$\pos$-RE\xspace}

\renewcommand{\phi}{\varphi}
\newcommand{\simmax}{\sim^m}

\newcommand{\prevS}{{\it prevS}}
\newcommand{\form}{{\it form}}
\newcommand{\simset}{{\it sim}}

\newcommand{\io}{
\SetKwInOut{Input}{input}\SetKwInOut{Output}{output}
\Input{a finite model $\gM=\tup{\Delta,\interp{\cdot}}$}
\Output{$F$, the set of \EL-formulas, and $S$, the simulator sets s.t.\
$(\forall v\in \Delta)\ \interp{F(v)}=S(v)=\simset_\EL(v)$}
\BlankLine}

\newcommand{\rg}{{\rm rg}}

\newcommand{\tup}[1]{\langle #1\rangle}
\newcommand{\cset}[1]{\{#1\}}

\newcommand{\gM}{\mathcal{M}}
\newcommand{\gG}{\mathcal{G}}
\newcommand{\gL}{\mathcal{L}}
\newcommand{\interp}[1]{|\!|#1|\!|}

\theoremstyle{plain}
\newtheorem{thm}{Theorem}

\theoremstyle{remark}
\newtheorem{ex}[thm]{Example}

\title{The Question of Expressiveness in the Generation of Referring Expressions}
\author{%
Carlos Areces\\
\small INRIA Nancy, Grand Est,\\
\small France\\
{\small\tt areces@loria.fr}
\And%
Santiago Figueira\\
\small Universidad de Buenos Aires,\\
\small Argentina\\
{\small\tt sfigueir@dc.uba.fr}
\And Daniel Gor\'in\\
\small Universidad de Buenos Aires,\\
\small Argentina\\
{\small\tt dgorin@dc.uba.fr}}

\begin{document}
\maketitle
\begin{abstract}
We study the problem of generating referring expressions modulo
different notions of expressive power.  We define the notion of
$\+L$-referring expression, for a formal language $\+L$ equipped
with a semantics in terms of relational models.  We show that the
approach is independent of the particular algorithm used to generate
the referring expression by providing examples using the frameworks
of~\cite{AKS08} and~\cite{Krahmer2003}. We provide some new
complexity bounds, discuss the issue of the length of the generated
descriptions, and propose ways in which the two approaches can be
combined.
\end{abstract}

\section{Generating referring expressions}

The generation of referring expressions (GRE) --given a context and
an element in that context generate a grammatically correct
expression in a given natural language that uniquely represents the
element-- is a basic task in natural language generation, and one of
active research
(see~\cite{dale89cooking,dale91:_gener_refer_expres_invol_relat,Dale1995,Stone2000,deemter02:_gener_refer_expres}
among others). Most of the work in this area is focused on the
\emph{content determination} problem (i.e., finding a collection of
properties that singles out the target object from the remaining
objects in the context) and leaves the actual \emph{realization}
(i.e., expressing a given content as a grammatically correct
expression) to standard techniques\footnote{For exceptions to this
practice see, e.g,~\cite{hora:algo97,ston:text98}}.

However, there is yet no general agreement on the basic
representation of both the input and the output to the problem; this
is handled in a rather ad-hoc way by each new proposal instead.

\cite{Krahmer2003} make the case for the use of \emph{labeled
directed graphs} in the context of this problem: graphs are abstract
enough to express a large number of domains and there are many
attractive and well-known algorithms for dealing with this type of
structures. Indeed, labeled directed graphs are nothing other than
an alternative representation of relational models, used to provide
semantics for formal languages like first and higher-order logics,
modal logics, etc. Even valuations, the basic models of
propositional logic, can be seen as one point labeled graphs. It is
not surprising then that they are well suited to the task.

In this article, we side with~\cite{Krahmer2003} and use labeled
graphs as input, but we argue that an important notion has been left
out when making this decision. Exactly because of their generality
graphs do not define, by themselves, a unique notion of
\emph{sameness}. When do we say that two nodes in the graphs can or
cannot be referred uniquely in terms of their properties?  This
question only makes sense once we fix a certain level of
expressiveness which determines when two graphs, or two elements in
the same graph, are equivalent.

Investigating the GRE problem in terms of different notions of
expressiveness is the main goal of this paper.  In
\S\ref{sec:technical}, we will show alternative (but equivalent)
ways in which different degrees of expressiveness can be defined,
and discuss how choosing the adequate expressiveness has an impact
on the number of instances of the GRE problem that have a solution
(less expressive logics can distinguish fewer instances); the
computational complexity of the GRE algorithms involved; and the
complexity of the surface realization problem.

We maintain that this perspective is independent of the particular
GRE algorithm being used. Our work fits naturally with the approach
of~\cite{AKS08} as we show in \S\ref{sec:simulation}, where we also
answer an open question concerning the complexity of the GRE problem
for the language \EL.  In \S\ref{sec:krahmer} we turn to the
subgraph based algorithm of~\cite{Krahmer2003}, and show how to
generalize it for other notions of sameness. In section
\S\ref{sec:combining} we show how one can combine the approaches of
the previous two sections and in section \S\ref{sec:size} we discuss
the size of the referring expressions relative to the expressiveness
employed.

\newcommand{\nDog}{\mathit{dog}\xspace}
\newcommand{\nCat}{\mathit{cat}\xspace}
\newcommand{\aSmall}{\mathit{small}\xspace}
\newcommand{\aSniffing}{\mathit{sniffs}\xspace}
\newcommand{\nBreed}{\mathit{beagle}\xspace}

\section{Measuring expressive power}\label{sec:technical}

Relational structures are notably appropriate for representing
\emph{situations} or \emph{scenes}.  A relational structure (also
called ``relational model'') is a non-empty set of objects --the
\emph{domain}-- together with a collection of relations, each with a
fixed arity.

Formally, assume a fixed and finite (but otherwise arbitrary)
vocabulary of $n$-ary relation symbols\footnote{We do not consider
constants or functions as they can be represented as relations of
adequate arity.}. Define a relational model $\+M$ as a tuple
$\tup{\Delta,\interp{\cdot}}$ where $\Delta$ is a nonempty set, and
$\interp{\cdot}$ is a suitable interpretation function, that is,
$\interp{r} \subseteq \Delta^n$ for every $n$-ary relation symbol
$r$. We say that $\+M$ is \emph{finite} whenever $\Delta$ is finite.
The \emph{size} of a model $\+M$ is the sum $\#\Delta +
\#\interp{\cdot}$, where $\#\Delta$ is the cardinality of $\Delta$
and $\#\interp{\cdot}$ is the sum of the sizes of all relations in
$\interp{\cdot}$.

\begin{figure}
 \begin{center}
 \begin{tikzpicture}
  [
    n/.style={circle,fill,draw,inner sep=1.5pt,node distance=1.5cm},
    aSniffing/.style={->, >=stealth, semithick, shorten <= 3pt, shorten >= 3pt},
  ]
 \node[n,label=above:$a$,label=below:{\relsize{-1}$\begin{array}{c}\nDog\end{array}$}] (a) {};
 \node[n,label=above:$b$,label=below:{\relsize{-1}$\begin{array}{c}\nDog\\ \aSmall \end{array}$}, right of=a] (b) {};
 \node[n,label=above:$c$,label=below:{\relsize{-1}$\begin{array}{c}\nCat\\ \aSmall\end{array}$}, right of=b] (c) {};
 \node[n,label=above left:$d$,label=below:{\relsize{-1}$\begin{array}{c}\nDog\\ \nBreed\\  \aSmall \end{array}$}, right of=c,xshift=-5pt] (d) {};
 \node[n,label=above right:$e$,,label=below:{\relsize{-1}$\begin{array}{c}\nCat\end{array}$},right of=d] (e) {};

 \draw [aSniffing,loop left] (a) to node[above,xshift=-5pt]{\relsize{-1}$\aSniffing$} (a);

 \draw [aSniffing,bend right=40] (b) to node[auto,swap]{\relsize{-1}$\aSniffing$} (a);

 \draw [aSniffing,bend right=40] (c) to node[auto,swap]{\relsize{-1}$\aSniffing$} (b);

 \draw[aSniffing, bend left=40] (d) to node[auto]{\relsize{-1}$\aSniffing$} (e);
 \draw[aSniffing, bend left=40] (e) to node[auto,swap]{\relsize{-1}$\aSniffing$} (d);

 \end{tikzpicture}
 \end{center}
 \vspace{-18pt}
 \caption{Graph representation of scene $\+S$.\label{fig:cat-dog-1}}
 \end{figure}
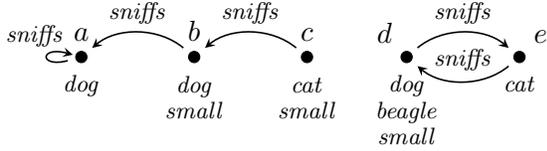

Figure~\ref{fig:cat-dog-1} shows the representation of a scene with
the relational model
 $\+S = \tup{\Delta,\interp{\cdot}}$:
 $$
 \begin{array}{c@{\,=\,}lc@{\,=\,}l}
\multicolumn{4}{c}{\Delta\,=\,\cset{a,b,c,d,e}}
\\
 \interp{\nDog} & \cset{a,b,d}
 &
\interp{\nCat} & \cset{c,e}
 \\
 \interp{\nBreed} & \cset{d}
& \interp{\aSmall} & \cset{b,c,d}
\\
\interp{\aSniffing} &
\multicolumn{3}{l}{\hspace{-5pt}\cset{(a,a),(b,a),(c,b),(d,e),(e,d)}}
 \end{array}
 $$

 Intuitively, $a$, $b$ and $d$ are dogs, while $c$ and $e$ are cats;  $d$ is a small beagle;
 $b$ and $c$ are also small.
 We read $\aSniffing(d,e)$ as ``$d$ is sniffing $e$''.

Logical languages are fitting for the task  of (formally)
\emph{describing} elements of a relational structure. Consider,
e.g., the classical language of first-order logic (with equality),
\FOL, given by:
$$
  \top \mid x_i \not\approx x_j \mid  r (\bar x) \mid \lnot \gamma \mid \gamma \land \gamma' \mid \exists x_i . \gamma
$$
where $\gamma,\gamma' \in \FOL$, $r$ is an $n$-ary relation symbol
and $\bar x$ is an $n$-tuple of variables. As usual, $\gamma \lor
\gamma'$ and $\forall x . \gamma$ are short for $\lnot(\lnot\gamma
\land \lnot\gamma')$ and $\lnot\exists x . \lnot\gamma$,
respectively. Formulas of the form $\top$, $x_i \not\approx x_j$ and
$r(\bar
x)$ are called \emph{atoms}.%
  \footnote{%
    Notice that we include the inequality symbol $\not \approx$ as
    primitive.  Equality can be defined using negation.
  }
Given a relational model $\+M = \tup{\Delta,\interp{\cdot}}$ and a
formula $\gamma$ with free variables%
\footnote{%
    W.l.o.g.\ we assume that no variable appears both free and bound, that no variable is bound
    twice, and that the index of bound variables in a formula increases from left to right.%
} among $x_1\ldots x_n$, we inductively define the \emph{extension}
or \emph{interpretation} of  $\gamma$ as the set of $n$-tuples
 $\interp{\gamma}^n \subseteq \Delta^n$ that satisfy:
 $$
\begin{array}{@{\hspace{-.5pt}}r@{\,=\,}l}
\interp{\top}^n & \Delta^n
\\
 \interp{x_i \not\approx x_j}^n & \cset{\bar{a} \mid \bar{a} \,{\in}\, \Delta^n, a_i \neq a_j}
 \\
\interp{r (x_{i_1} \ldots x_{i_k})}^n & \cset{\bar{a} \mid \bar{a}
\,{\in}\, \Delta^n, (a_{i_1} \ldots a_{i_k}) {\in} \interp{r}}
\\
  \interp{\lnot\delta}^n & \Delta^n \setminus \interp{\delta}^n
\\
  \interp{\delta \land \theta}^n & \interp{\delta}^n \cap \interp{\theta}^n
\\
  \interp{\exists x_{l}.\delta}^n & \cset{\bar a \mid \bar a  a_{n+1}  \in \interp{\delta'}^{n+1}}
\end{array}
$$
where $1 \le i,j, i_1, \ldots, i_k \le n$, $\bar{a} = (a_1\ldots
a_n)$, $\bar{a}a_{n+1} = (a_1\ldots a_{n+1})$ and $\delta'$ is
obtained by replacing all occurrences of $x_l$ in $\delta$ by
$x_{n+1}$. When the cardinality of the tuples involved is known from
context we will just write $\interp{\gamma}$ instead of
$\interp{\gamma}^n$.

With a language syntax and semantics in place, we can now formally
define the problem of $\+L$-GRE for a target set of elements $T$\footnote{%
  Similarly we can use formulas with two free variables to describe binary relations, etc.
  Throughout this paper though, we only discuss the GRE for single elements.%
} (we slightly adapt the definition in~\cite{AKS08}):

\medskip
\noindent {\small
\begin{center}
\begin{tabular}{ll} \hline
\multicolumn{2}{l}{ \textsc{$\gL$-GRE Problem}}\\ \hline
\ \ Input: & a model $\gM=\tup{\Delta,\interp{\cdot}}$ and a target\\
&  set $T \subseteq \Delta$.\\
\ \ Output: & a formula $\varphi \in \gL$ such that
$\interp{\varphi} = T$, or\\
& $\bot$ if such a formula does not exists.\\ \hline
\end{tabular}
\end{center}}
In case the output is not $\bot$, we say that $\phi$ is a
$\+L$-referring expression ($\+L$-RE) for $\+M$.

\subsection{Choosing the appropriate language}\label{sec:choosinglanguage}

Given a model $\+M$, there will be an infinite number of formulas
that uniquely describe a target (e.g., if $\varphi$ describes a
target $T$, then $\varphi \wedge \varphi$ trivially describes $T$ as
well; even formulas which are not logically equivalent might have
the same interpretation once the model is fixed). Despite having the
same interpretation in $\+M$, they may be quite different with
respect to other parameters.

To start with, and as it is well known in the automated text
generation community, different realizations of the same content
might result in expressions more or less appropriate. Although, as
we mentioned in the introduction, we will only address the content
determination (and not the surface realization) part of the GRE
problem, generating content using languages with different
expressive power can have an impact in the posterior surface
generation step.

Let us consider again the scene in Figure~\ref{fig:cat-dog-1}.
Formulas $\gamma_1$--$\gamma_4$ shown in Table~\ref{tab:gammas} are
all such that $\gamma_i$
 uniquely describe $b$ in model $\+S$.

\begin{table}
\begin{small}
$$
\begin{array}{cl}
 \gamma_1: & \nDog(x) \land \aSmall(x) \land
   \exists y . (\aSniffing(x,y) \land \nDog(y))\\[3pt]
  \gamma_2: & \nDog(x) \land \aSmall(x) \land
  \forall y . (\neg \nCat(y) \lor \neg \aSniffing(x,y))\\[3pt]
  \gamma_3: & \nDog(x) \land
  \exists y . (x \not\approx y \land \nDog(y)  \land \aSniffing(x,y))\\[3pt]
  \gamma_4: & \nDog(x) \land
  \exists y . (\nCat(y) \land \aSmall(y) \land \aSniffing(y,x))
 \end{array}
$$
\end{small}
\vspace{-18pt} \caption{Descriptions for $b$ in
Figure~\ref{fig:cat-dog-1}.\label{tab:gammas}}
\end{table}

Arguably, $\gamma_1$ can be easily realized as \emph{``the small dog
that sniffs a dog''}. Syntactically, $\gamma_1$ is characterized as
a positive, conjunctive, existential formula (i.e., it contains no
negation and uses only conjunction and existential quantification).
Expressions with these characteristics are, by large, the most
commonly found in corpora as those compiled
in~\cite{viet:algo06,deem:buil06,dale:refe09}. $\gamma_2$ on the
other hand contains negation, disjunction and universal
quantification and could be realized as \emph{``the small dog that
only sniffs things that are not cats''} which sounds very unnatural.
Even a small change in the form of $\gamma_2$ turns it more
palatable: rewrite it using only $\exists$, $\lnot$, and $\land$ to
obtain \emph{``the small dog that is not sniffing a cat''}.
Similarly, formulas $\gamma_3$ and $\gamma_4$ seem computationally
harder to realize than $\gamma_1$: $\gamma_3$ because it contains an
inequality (\emph{``the dog sniffing another dog''}) and $\gamma_4$
because the quantified object appears in the first argument
possition in the binary relation (\emph{``the dog that is sniffed by
a small cat''}).

Summing up, even without taking into account fundamental linguistics
aspects that will make certain realization preferable --e.g.,
saliency, the cognitive capacity of the hearer (can she recognize a
\emph{beagle} from another kind of dog?), etc.-- we can ensure
during content determination certain properties of the generated
referring expression.

%the content determination problem can be seen as that of generating a certain logical
%formula, but not just any formula. It is not simple to characterize which formulas are not acceptable
%but one can approximate it by restricting to formulas of certain shape.

Concretely, let \EPFOL be the fragment of \FOL-formulas where the
operator $\lnot$
does not occur.%
\footnote{%
  But notice that atoms $x_i \not\approx x_j$ are permitted.
} By restricting content determination to \EPFOL, we ensure that
formulas like  $\gamma_2$ will not be generated. If we also (or,
alternatively) ban $\not\approx$ from the language, $\gamma_3$ is
precluded. And we need not restrict ourselves to \emph{explicit}
fragments of first-order logic: many logical languages are known to
be expressively equivalent to fragments of first-order logic.  For
example, the language of the description logic
\ALC~\cite{baad:desc03}, given by:
$$
\top \mid p \mid \neg \gamma \mid \gamma \wedge \gamma' \mid
\exists r. \gamma
$$
(where $\gamma,\gamma' \in \ALC$) corresponds to a syntactic
fragment of \FOL without $\not\approx$, as shown by the standard
translation to first-order logic $\st_x$:
\begin{align*}
 \st_{x_i}(\top) &= \top
\\
  \st_{x_i}(p) &= p(x_i)
\\
 \st_{x_i}(\lnot \gamma) &= \lnot\st_{x_i}(\gamma)
\\
 \st_{x_i}(\gamma_1 \land \gamma_2) &= \st_{x_i}(\gamma_1) \land \st_{x_i}(\gamma_2)
\\
 \st_{x_i}(\exists r . \gamma) &= \exists x_{i+1} . (r(x_i,x_{i+1}) \land \st_{x_{i+1}}(\gamma))
\end{align*}
%
%
% $$
% \begin{array}{c@{\,=\,}lr@{\,=\,}l}
% \st_{x_i}(\top) & \top & \st_{x_i}(\gamma_1\! \land\! \gamma_2) & \st_{x_i}(\gamma_1)\! \land\! \st_{x_i}(\gamma_2)
% \\
%\st_{x_i}(p) & p(x_i) &  \st_{x_i}(\lnot \gamma) &
%\lnot\st_{x_i}(\gamma)
%\\
%\multicolumn{4}{c}{\st_{x_i}(\exists r . \gamma) \,=\, \exists
%x_{i+1} . (r(x_i,x_{i+1})\! \land\! \st_{x_{i+1}}(\gamma))}
% \end{array}
% $$
%
Hence, by restricting content generation to $\ALC$ we would avoid
formulas like $\gamma_3$ (no equality) and $\gamma_4$ (quantified
element appears always in second argument position).

\cite{AKS08} discuss generation in terms of different description
logics like \ALC and \EL (\ALC without negation). In this article,
we will extend the results in that paper, considering for instance
\ELAN (\ALC with negation allowed only in front of unary relations)
but, more generally, we argue that the primary question is not
whether one should use one or other (description) logic for content
generation, but rather which are the \emph{semantic differences} one
cares about. This determines the required logical formalism which,
in turn, impacts both content determination and surface realization.

%\emph{level of
%expressiveness} we require.  As we discussed earlier the syntactic form
%of the generated formula does have an impact in the GRE problem (as we
%show with formula $\gamma_2$, a logically equivalent formula might be
%easier to realize). We will return to this issue in \S\ref{sec:size}.
%But a more fundamental issue is at hand.
%Before we worry about the \emph{shape} of the formulas we generate, we
%should decide which are the \emph{semantic differences} we care
%about, and these are determined by the intrinsic expressiveness of the
%chosen logical formalism
%\fixme{C: agregue esto ultimo, vean que les parece. d: me parece bien!}

%\fixme{Santi: tal vez cambiar el ejemplo de las $\gamma$ para que
%alguna sea \ELAN?} Notice that $\gamma_1$--$\gamma_3$ do not
%correspond to \EL-formulas, but $\gamma_4$ does.
%\fixme{Footenote
%needed? CARLOS} \footnote{
%  Needless to say, it is possible to generate dubious \EL-formulas that uniquely describe %$a$, like
%  $\exists y . (\nDog(y) \land \aSniffing(x,y) \land \exists z . (\nDog(z) \land %\aSmall(z) \land \aSniffing(y,z)))$.
%}

We have mentioned several logic languages (and there are many more
alternatives like allowing disjunctions, counting quantifiers,
etc.). Each language can be seen as a compromise between
expressiveness, realizability and computational complexity.
Therefore, the appropriate selection for a particular GRE task
should depend on the actual context. Moreover, as we will see, the
move from one logical language to another impacts not only on the
shape of formulas that can be generated but also on the
computational complexity of the generation problem, and on its
success, i.e., when it will be possible to uniquely identify a given
target.

\subsection{Defining \emph{sameness}}

For any given logical language $\+L$,  we say that $u$ is
distinguishable (in $\+L$) from $v$ whenever there exists an
$\+L$-formula $\gamma$ such that $u$ makes $\gamma$ true while $v$
makes it false. Formally,
%let $\+L$ stand for any of the languages discussed so far, and
let $\+M_1 = \tup{\Delta_1, \interp{\cdot}_1}$ and $\+M_2 =
\tup{\Delta_2, \interp{\cdot}_2}$ be two relational models with $u
\in \Delta_1$ and $v \in \Delta_2$; we follow the terminology
of~\cite{AKS08} and say that ``$u$ is $\+L$-similar to $v$''
(notation $u \simil{\+L} v$) whenever $u \in \interp{\gamma}_1$
implies
$v \in \interp{\gamma}_2$, for every $\gamma \in \+L$.%
\footnote{%
   For $\gamma \in \+\ALC$ and its sublanguages, $\interp{\gamma}=\interp{\st_{x_1}(\gamma)}$.
  }
$\+L$-similarity is reflexive for all $\+L$, and symmetric  for
languages that contain negation.

Observe that $\+L$-similarity captures the notion of
`indistinguishability' (in $\+L$). One can take $\+M_1$ and $\+M_2$
to be the same model and in that case, if $u \simil{\+L} v$ for $u
\neq v$, the $\+L$-content determination problem
%\fixme{C: decia ``may never succeed'' lo cambie a ``will never succeed''}
for $u$ will not succeed (since for every $\gamma \in \+L$,
$\interp{\gamma}\neq\cset{u}$).

Fortunately, one need not consider infinitely many $\+L$-formulas to
decide whether $u$ is $\+L$-similar to $v$. We can reinterpret
$\+L$-similarity in terms of standard model-theoretic notions like
isomorphisms or
bisimulations\footnote{%
For the rest of the article, we will focus on relational models with
only unary and binary relational symbols. These are the usual models
of interest when describing scenes as the one presented in
Figure~\ref{fig:cat-dog-1}. Accommodating relations of higher arity
poses only notational problems. } which describe structural
properties of the model, instead. We will use the term
\emph{$\+L$-simulation} to refer to the suitable notion for $\+L$;
in the case of the languages we are considering they can be defined
in a modular way. Given a relation $\sim \subseteq \Delta_1 \times
\Delta_2$, it may or may not possess the properties we call
$\atomLR$, $\zigzag$, $\injLR$ (see below), Table~\ref{tab:simu}
define various $\+L$-simulations in terms of these.

%\begin{description}
%\item[\smaller$\atomL$:] If $u_1{\sim} u_2$, then $u_1 \in \interp{p}_1 \Rightarrow u_2 \in \interp{p}_2$.
%\item[\smaller$\atomR$:] If $u_1{\sim} u_2$, then $u_2 \in \interp{p}_2 \Rightarrow u_1 \in \interp{p}_1$.
%\item[\smaller$\zig$:] If $u_1{\sim} u_2$ and $(u_1,v_1) \in \interp{p}_1$, then $v_1{\sim}v_2$
%  and $(u_2,v_2) \in \interp{p}_2$, for some $v_2$.
%\item[\smaller$\zag$:] If $u_1{\sim}u_2$ and $(u_2,v_2) \in \interp{p}_2$, then $u_1{\sim}v_1$ and
% $(u_1,v_1) \in \interp{p}_1$, for some $v_1$.
%\item[\smaller$\injL$:] $\sim$ is an injective function $\Delta_1 \to \Delta_2$.
%\item[\small$\injR$:] $\sim^{-1}$ is an injective function $\Delta_2 \to \Delta_2$.
%\end{description}

\newcommand{\simdef}[2]{\noindent\textbf{\smaller$#1$:}\hfill\parbox[t]{.4\textwidth}{#2}\par}

\simdef{\atomL}{If $u_1{\sim} u_2$, then $u_1 \in \interp{p}_1
\Rightarrow u_2 \in \interp{p}_2$.} \simdef{\atomR}{If $u_1{\sim}
u_2$, then $u_2 \in \interp{p}_2 \Rightarrow u_1 \in \interp{p}_1$.}
\simdef{\zig}{If $u_1{\sim} u_2$ and $(u_1,v_1) \in \interp{p}_1$,
then $v_1{\sim}v_2$
  and $(u_2,v_2) \in \interp{p}_2$, for some $v_2$.}
\simdef{\zag}{If $u_1{\sim}u_2$ and $(u_2,v_2) \in \interp{p}_2$,
then $u_1{\sim}v_1$ and
 $(u_1,v_1) \in \interp{p}_1$, for some $v_1$.}
\simdef{\injL}{$\sim:\Delta_1 \to \Delta_2$ is an injective
function.} \simdef{\injR}{$\sim^{-1}:\Delta_2 \to \Delta_2$ is an
injective function.}

The following is a fundamental model-theoretic
result~\cite{ebbi:math96,KR99,BRV01}:
\begin{thm} \label{thm:simulation}
If  $\+M_1$ and $\+M_2$ are finite models, $u \in \Delta_1$ and $v
\in \Delta_2$, then $u \simil{\+L} v$ iff $u \simul{\+L} v$.
\end{thm}
The right to left implication does not hold in general on infinite
models. Notice that $\simul{\FOL}$ corresponds to relational model
isomorphism while $\simul{\ALC}$ corresponds to the notion of
bisimulation.

$\+L$-simulations allow us to determine, in an effective way, when
an object is indistinguishable from another in a given model with
respect to $\+L$.

For example, it is easy to see that $a \simul{\EL} b$ in the model
of Figure~\ref{fig:cat-dog-1} (simply verify that the $\sim$ such
that $a \sim b$ and $x \sim x$ (for $x \in \Delta$) satisfies
$\atomL$ and $\zig$). Using Theorem~\ref{thm:simulation} we conclude
that, intuitively, no \EL-formula can distinguish ``a dog sniffing
itself'' from ``a dog sniffing (another) dog sniffing itself''.
Similarly, no $\FOL$ formula will distinguish two $\FOL$-similar
(isomorphic) objects.

\begin{table}
\newcommand{\colsep}{\hspace{6pt}}
\begin{flushleft}
\begin{smaller}
$$
\begin{array}{c|c@{\colsep}c@{\colsep}c@{\colsep}c@{\colsep}c@{\colsep}c}
  \+L & \multicolumn{6}{l}{u \simul{\+L} v \text{ iff } u\sim v \text{ for a relation $\sim$ that satisfies:}}\\
  \hline
  \FOL & \atomL & \atomR & \zig & \zag & \injL & \injR\\
  \EPFOL & \atomL && \zig && \injL &\\
  \ALC & \atomL & \atomR & \zig & \zag&&\\
  \EL   & \atomL &&  \zig & &\\
  \ELAN  & \atomL &\atomR&  \zig & &\\
\end{array}
$$
\end{smaller}
\end{flushleft}
\vspace*{-18pt} \caption{Simulations for various
logics.}\label{tab:simu}
\end{table}

There are well-known algorithms to compute certain kinds of
$\+L$-simulations~\cite{H71,PT87,HHK95,DPP03}. We will discuss this
in more detail in the next section.  While polynomial time
algorithms for many languages like \ALC, \ALC with inverse
relations,  \ELAN and \EL-simulation, etc.\ can be obtained, no
polynomial time algorithms for \FOL and \EPFOL-simulation are known
(actually, even their actual complexity class is not
known~\cite{gare:comp79}).

\section{GRE via simulator sets}\label{sec:simulation}

Given a model $\+M = \tup{\Delta, \interp{\cdot}}$,
Theorem~\ref{thm:simulation} tells us that if two distinct elements
$u$ and $v$ in $\Delta$ are such that $u \simul{\+L} v$ then every
$\+L$-formula that is true at $u$ is also true at $v$. Hence there
is no formula in $\+L$ that can uniquely refer to $u$. From this
perspective, knowing whether the model contains an element that is
$\gL$-similar but distinct from $u$ is equivalent to that of whether
there exists an $\+L$-RE for $u$.

We begin by considering this task. Assume fixed a language $\+L$ and
a model $\+M$.  Suppose we want to refer to an element $u$ in the
domain of $\+M$. We would like to compute the {\em simulator set} of
$u$ defined as $\simset_{\+L}(v) = \cset{u \mid v \simul{\+L} u}$.
 %of all elements in the domain which are $\+L$-similar to $u$.
 If $\simset_{\+L}(v)$ is not the singleton $\cset{v}$,
% contains elements beside $v$ then
 we cannot $\+L$-refer to $v$.
\iffullversion It is easy to see that the union of two
$\+L$-simulations is also an $\+L$-simulation. We can then define
the \emph{maximal auto $\+L$-simulation} (notation, $\simmax_{\+L}$)
over a model $\+M$ as the union of all auto $\+L$-simulations over
$\+M$. Because $\simset_{\+L}(u) = \cset{v \mid u \simmax_{\+L} v}$,
an algorithm for computing $\simmax_{\+L}$ also computes
$\simset_{\+L}(u)$. \else \fi

%$G=(N,\to,l)$, notated
%$a\simmax_P b$ is defined as follows: for any $a,b\in N$,
%$a\simmax_P b$ iff there is a $P$-simulation $\sim$ over $G$ such
%that $a\sim b$. It is not difficult to see that $\simmax_P$ is
%indeed the maximal $P$-simulation, i.e., $\simmax_P$ is a
%$P$-simulation and for any $P$-simulation $\sim$, we have
%$\simmax_P\supseteq\sim$. We write $\simmax$ for $\simmax_{\Id}$.

\iffullversion If $P$ is reflexive and transitive then so is
$\simmax_P$. In particular, $\simmax$ is reflexive and transitive.
\fixme{Are reflexivity and transitivity important? Check.} \else \fi

%Given a labeled graph $G=(N,\to,l)$ and given $P\subseteq(\rg\
%l)^2$, let $\simset_P(v)\subseteq N$ denote the set of nodes which
%$P$-simulate $v$, i.e.
%$$
%\simset_P(v)=\{u\in N \colon v\simmax u\}.
%$$
%$\simset_P(v)$ is called the \emph{$P$-simulator set of $v$}. We
%write $\simset$ for $\simset_{\Id}$.

%Given a graph $\gG$ and $v$ a node,
%the \emph{$P$-simulator set of $v$} is the set
%$\simset_P(v) = \{w : v \simmax_P w\}$.
%$P$-simulate $v$, i.e.
%$$
%\simset_P(v)=\{u\in N \colon v\simmax u\}.
%$$
%$\simset_P(v)$ is called . We
%write $\simset$ for $\simset_{\Id}$.

%Given a relation $P\subseteq (\rg \ l)^2$ and a graph $G=(N,\to,l)$,
%we aim at computing the maximal $P$-simulation of $G$.

\cite{HHK95} propose an algorithm to compute the set
$\simset_{\ELAN}(v)$ for each element $v$ of a given finite model
$\+M=\tup{\Delta,\interp{\cdot}}$\footnote{%
  Actually the algorithm proposed in \cite{HHK95} is over labeled graphs, but
  it can be adapted to compute $\simset_{\ELAN}$ by
  appropriately labeling the model.%
} in time $O(\size{\Delta}\times\size{{\interp{\cdot}}})$.
Intuitively, this algorithm defines $S(v)$ as the set of candidates
for simulating $v$ and successively refines it by removing those
elements which do not simulate $v$.
%Since we never put new vertices into $\simset(v)$, all the deletions
%from $\simset(v)$ are permanent.
At the end, $S(v)=\simset_\ELAN(v)$.

The algorithm can be adapted to compute $\simset_\+L$ for other
$\+L$. In particular, we can use it to compute $\simset_\EL$ in
polynomial time which will give us the basic algorithm for
establishing an upper bound to the complexity of the \EL-GRE problem
(this will answer an open question of~\cite{AKS08}).

Let us first introduce some notation. We fix $\+P$ as the set of all
unary relations of the language \EL. For $v\in \Delta$ let
$\unary(v)=\{p\in\+P\mid v\in\interp{p}\}$ and let
$\su{r}{v}=\{u\in\Delta\mid(v,u)\in\interp{r}\}$ for any binary
relation $r$ present in the language.
%
%
%Let us first introduce some notation. For $v\in \Delta$ let
%$\unary(v)=\{p\mid \mbox{$p$ is a unary relation and
%$v\in\interp{p}$}\}$ and let
%$\su{r}{v}=\{u\in\Delta\mid(v,u)\in\interp{r}\}$ for any binary
%relation $r$.
%\begin{eqnarray*}
%\unary(v)&=&\{p\mid \mbox{$p$ is a unary rel.\ and
%$v\in\interp{p}$}\}\\
%%\pr{r}{v}&=&\{u\in\Delta\mid(u,v)\in\interp{r}\}\\
%\su{r}{v}&=&\{u\in\Delta\mid(v,u)\in\interp{r}\}
%\end{eqnarray*}

%First some notation. For any $v\in \Delta$ and any binary relation
%$r$, let
%\begin{eqnarray*}
%\unary(v)&=&\{p\mid \mbox{$p$ is a unary rel.\ and
%$v\in\interp{p}$}\}\\
%%\pr{r}{v}&=&\{u\in\Delta\mid(u,v)\in\interp{r}\}\\
%\su{r}{v}&=&\{u\in\Delta\mid(v,u)\in\interp{r}\}
%\end{eqnarray*}
%%The last two extend to sets $V\subseteq\Delta$ as usual:
%%$\pr{r}{V}=\bigcup_{v\in V}\pr{r}{v}$ and $\su{r}{V}=\bigcup_{v\in
%%V}\su{r}{v}$.

%
The pseudo-code is shown in Algorithm~\ref{alg:schematic-gen-sim}.
We initialize $S(v)$ with the set of all elements $u\in\Delta$ such
that $\unary(v)\subseteq \unary(u)$, i.e., the set of all elements
satisfying at least the same unary relations as $v$ (this guarantees
that $\atomL$ holds). At each step, if there are three elements $u$,
$v$ and $w$ such that for some relation $r$, $(u,v) \in \interp{r}$,
$w\in S(u)$ (i.e., $w$ is a candidate to simulate $u$) but
$\su{r}{w}\cap S(v)$ (there is no element $w'$ such that $(w, w')
\in \interp{r}$ and $w'\in S(v)$) then clearly condition $\zig$ is
not satisfied under the supposition that $\simset_\EL=S$. $S$ is
`too big' because $w$ cannot simulate $u$. Hence $w$ is removed from
$S(u)$.
%This
%intuitive algorithm does not lead to an
%$O(\size{\Delta}\times\size{{\interp{\cdot}}})$\fixme{Santi: ver bien
%la complejidad.} running time, but an efficient refinement does.

%The algorithm can be easily adapted to calculate the maximal
%$P$-simulation, for a given binary relation $P$. Since the only
%difference between a simulation and a $P$-simulation is the
%relation used in condition~\ref{item:gsim:1} of Definitions~\ref{def:gsim},
%we only need to rearrange the initialization of the
%set $S(v)$ for each node $v$ by setting $S(v)=\{u\in N\colon l(v)\ P\ l(u)\}$.

%Algorithm~\ref{alg:schematic-gen-sim} gives a schematic view of the
%process of calculating $\simset_P(v)$ for each vertex $v$.

\begin{algorithm} \small
\caption{\small Computing
\EL-similarity}\label{alg:schematic-gen-sim}
\SetKwInOut{Input}{input}\SetKwInOut{Output}{output}
\Input{a finite model $\+M=\tup{\Delta,\interp{\cdot}}$}
\Output{$\forall v\in \Delta$, the simulator set
$\simset_\EL(v)=S(v)$} \BlankLine

\ForEach{$v\in \Delta$}{$S(v):=\{u\in \Delta \mid \unary(v)\
\subseteq \unary(u) \}$}

\While{\guard}{$S(u):=S(u)\setminus\{w\}$}
\end{algorithm}

Of course, Algorithm~\ref{alg:schematic-gen-sim} will only tell us
whether a referring expression for an element $v$ exists (that is,
whenever $\simset_{\EL}(v) = \cset{v}$).  It does not compute an
\EL-formula $\varphi$ that uniquely refers to $v$. But
Algorithm~\ref{alg:schematic-gen-sim} is an instance of a family of
well-known algorithms that compute $\+L$-simulations by successively
refining an over-approximation of the simulator sets. The ``reason''
behind each refinement  can be encoded using an $\+L$-formula;
intuitively, nodes that do not satisfy it are being removed from the
simulator set on each refinement.

Using this insight, one can transform an algorithm that computes
$\+L$-simulator sets into one that additionally computes an $\+L$-RE
for each set. \cite{AKS08} used this approach to derive their
\ALC-GRE method from a well-known algorithm for computing
\ALC-simulation (i.e., bisimulation) sets, but failed to notice they
could derive one for \EL analogously.

Algorithm~\ref{alg:schematic-GRE} shows a transformed version of
Algorithm~\ref{alg:schematic-gen-sim} following this principle. The
idea is that each node $v\in\Delta$ is now tagged with a formula
$F(v)$ of \EL. The formulas $F(v)$ are updated along the execution
of the loop, whose invariant  ensures that $v \in \interp{F(v)}$ and
$\interp{F(u)} \subseteq S(u)$ hold for all $u,v\in\Delta$.

\begin{algorithm}\small
%\LinesNumbered
\io

\ForEach{$v\in \Delta$}{ $S(v):=\{u\in \Delta \mid \unary(v)\
\subseteq \unary(u) \}$\label{alg:line:init1}

$F(v):=\bigwedge \unary(v)$\label{alg:line:init2} }

\While{\guard} {
 %$\{\ I: \mbox{\bf assert }(\forall u,v)\ \interp{F(u)} \subseteq S(u)\wedge v \in \interp{F(v)} \ \}$
 \KwSty{invariant} $(\forall u,v)\ \interp{F(u)} \subseteq S(u)\wedge v \in \interp{F(v)}$\;
$S(u):=S(u)\setminus\{w\}$\label{alg:line:loop-body-begin}

\If{$\diam F(v)$ is not a conjunct of $F(u)$}{ $F(u):=F(u)\wedge
\diam F(v)$\label{alg:line:loop-body-end-1} }} \caption{\small
Computing $\EL$-similarity and \posre}\label{alg:schematic-GRE}
\end{algorithm}

Initially $F(v)$ is the conjunction of all the unary relations that
satisfy $v$ (if there is none, then $F(v)=\top$). Next, each time
the algorithm finds $r,u,v,w$ such that $(u,v)\in\interp{r}$, $w\in
S(u)$ and $\su{r}{w}\cap S(v)=\emptyset$, it updates $F(u)$ to
$F(u)\wedge\diam F(v)$. Again this new formula $\phi$ is in $\pos$
and it can be shown that $v\in\interp{\phi}$ and
$w\notin\interp{\phi}$, hence witnessing that $v\simil{\EL} w$ is
false.

%As we will see in Theorem~\ref{thm:correctness-schematic-GRE}, this
%formula is true in $v$ but false in $w$, hence witnessing that $w$
%does not simulate $v$.

%It is clear that Algorithm \ref{alg:schematic-GRE} terminates.

\iffullversion One wants to know whether a given set $V\subseteq W$
has an \posre. How do we interpret the output of Algorithm
\ref{alg:schematic-GRE}? Here is the answer: $V\subseteq W$ has an
\EL-RE iff there is a node $v\in V$ such that $V=\{u\in W\colon
v\in\simset_\subseteq(u) \wedge u\in\simset_\subseteq(v)\}$. In
other words, $V$ has an \EL-RE iff $V$ is the set of nodes which are
\emph{bisimilar} to some node of $V$. Here we say that $u$ and $v$
are \emph{bisimilar} if $u\in\simset_\subseteq(v)$ and
$v\in\simset_\subseteq(u)$. In case $V=\{v_1,\dots,v_n\}$ has an
\posre, any $F(v_i)$ is a valid \posre, so one can pick any of them.

In particular, $v$ has an \EL-RE iff $\simset_\subseteq(v)=\{v\}$
and in case $v$ has a \EL-referring expression then $F(v)$ is a
valid one. If $v$ does not have an \posre then the algorithm may be
used to approximate an \posre. Indeed, since every formula true at
$v$ is also true at all nodes in $\simset_\subseteq(v)$ and $F(v)$
is true at every node of $\simset_\subseteq(v)$ then $F(v)$ is a
reasonable approximation of an \EL-RE for $v$.

\begin{ex}
Let $\gG$ be the following model
\begin{center}
\begin{tikzpicture}[>=latex]
  \node (n1) at (0,0) [shape=circle,draw,inner sep=2pt, label=right:$$] {$1$} ;
  \node (n2) at (2,0) [shape=circle,draw,inner sep=2pt, label=right:$$] {$2$} ;
  \node (n4) at (0,1) [shape=circle,draw,inner sep=2pt, label=right:$p$] {$3$} ;
  \node (n5) at (2,1) [shape=circle,draw,inner sep=2pt, label=right:$p{,}q$] {$4$} ;
  \draw [->] (n1) -- (n4);
  \draw [->] (n2) -- (n5);
\end{tikzpicture}
\end{center}
where $\rg\ l=\cset{p,q}$ and the valuation is defined as
$l(3)=\cset{p}, l(4)=\cset{p,q}, l(1)=l(2)=\cset{}$. The initial
values of $S$ and $F$ are shown in Table~\ref{tab:example}.

\begin{table}[ht]
\centering {\footnotesize
\begin{tabular}{|c|c|c|c|c|}
\hline
&\multicolumn{2}{|c|}{Initial}&\multicolumn{2}{|c|}{Final}\\
$v$ & $S(v)$ & $F(v)$ & $S(v)$&$F(v)$\\
\hline
$1$ & $\{1,2,3,4\}$ & $\top$ & $\{1,2\}$&$\top\wedge\diam p$\\
$2$ & $\{1,2,3,4\}$ & $\top$ & $\{2\}$&$\top\wedge\diam(p\wedge q)$\\
$3$ & $\{3,4\}$ & $p$ & $\{3,4\}$ &$p$\\
$4$ & $\{4\}$ & $p\wedge q$ & $\{4\}$&$p\wedge q$\\
\hline
\end{tabular}
\caption{Initial and final values of $F$ and $S$}\label{tab:example}
}
\end{table}

Suppose the following execution:
\begin{enumerate}
\item Choose $u=2,v=4,w=1$: detect that $1$ does not simulate $2$; set $S(2)=\{2,3,4\}$ and $F(2)=\top\wedge\diam(p\wedge q)$
\item Choose $u=2,v=4,w=3$: detect that $3$ does not simulate $2$; set $S(2)=\{2,4\}$ and $F(2)=\top\wedge\diam(p\wedge q)$
\item Choose $u=2,v=4,w=4$: detect that $4$ does not simulate $2$; set $S(2)=\{2\}$ and $F(2)=\top\wedge\diam(p\wedge q)$
\item Choose $u=1,v=3,w=3$: detect that $3$ does not simulate $1$; set $S(1)=\{1,2,4\}$ and $F(1)=\top\wedge\diam p$
\item Choose $u=1,v=3,w=4$: detect that $4$ does not simulate $1$; set $S(1)=\{1,2\}$ and $F(1)=\top\wedge\diam p$
\end{enumerate}
After the fifth iteration it terminates. The final output is shown
in Table~\ref{tab:example}. From this output one may conclude that,
since $\simset_\subseteq(4)=\{4\}$, node $4$ has an \posre, namely
$p\wedge q$. Node $2$ also has the \EL-RE $\top\wedge\diam(p\wedge
q)$. In contrast $3$ does not have an \EL-RE because every
$\pos$-formula true at $3$ is also true at $4$ (in this case, the
only such possible formula is $p$ itself, or logically equivalent
formulas such as $\top\wedge p\wedge p$). Nor $3$ has an \posre.
\end{ex}
\fi

\iffullversion
\begin{thm}\label{thm:correctness-schematic-GRE}
Let $S$ and $F$ be the output of the Algorithm
\ref{alg:schematic-GRE} with input $\gG=\tup{N,\to,l}$. Then for
each node $v\in N$, $\interp{F(v)} = S(v) = \simset_\subseteq(v)$
\end{thm}
\fi

\iffullversion
\begin{proof}
It is clear that for each node $v\in N$,
$\simset_\subseteq(v)=S(v)$. For the second part, we propose the
following invariant for the main loop:
\begin{itemize}
\item[$I_1$:] for each $v\in N$, $v \in \interp{F(v)}$
\item[$I_2$:] for each $u,v\in N$, $\interp{F(u)} \subseteq S(u)$
\end{itemize}
Let us analyze the state after the initialization, before line
\ref{alg:line:loop-begin} is executed for the fist time. On the one
hand, it is clear that for any node $v\in N$, $v \in \interp{F(v)}$,
so $I_1$ is verified. On the other, if $v\notin S(u)$ then
$l(u)\not\subseteq l(v)$ and so there is a propositional symbol $p$
such that $p \in l(u)$ and $p \notin l(v)$. Since
$v\notin\interp{p}$ then $v\notin\interp{\bigwedge l(u)}$ and
therefore $v\notin \interp{F(u)}$.

Suppose that $I_1$ and $I_2$ are true before executing
line~\ref{alg:line:loop-body-begin}. For all $v\in N$ let $S(v)=S_v$
and $F(v)=\phi_v$ in this state. Let $u,v,w$ be the chosen nodes. We
show that the invariant is reestablished after executing line
\ref{alg:line:loop-body-end-1}. Since $F$ and $S$ only change for
$u$, it suffices to show
\begin{itemize}
\item[$I_1'$:] $u \in \interp{\phi_u\wedge\diam \phi_v}$
\item[$I_2'$:] $\forall v\in N, \interp{\phi_u\wedge\diam \phi_v} \subseteq S_u\setminus\{w\}$
\end{itemize}
For $I_1'$, it is clear from $I_1$ that $u \in \interp{\phi_u}$ and
$v \in \interp{\phi_v}$. Since $u\to v$ we conclude $u \in
\interp{\diam \phi_v}$.

For $I_2'$ we show that for any $v$, if $v\notin S_u\setminus\{w\}$
then $v\notin \interp{\phi_u\wedge\diam \phi_v}$. If $v\notin S_u$,
by $I_2$ we know $v\notin \interp{\phi_u}$ and therefore $v\notin
\interp{\phi_u\wedge\diam \phi_v}$. Suppose $v=w$ and, by
contradiction, assume $v \in \interp{\phi_u\wedge\diam \phi_v}$.
Then there is $w'\in N$ such that $v\to w'$ and $w' \in
\interp{\phi_v}$. By $I_2$ this implies $w'\in S_v$. Hence
$w'\in\su(w)\cap S_v$ and this contradicts the choice of $u,v,w$.

When the algorithm terminates, $S=\simset_\subseteq$ and therefore
the invariant $I_2$ implies that for each $u,v\in N$, $F(u)
\subseteq \simset_\subseteq(u)$. On the other hand, $I_1$ implies
that $u \in \interp{F(u)}$ and by Proposition \ref{prop:equiv} we
have that if $v\in \simset_\subseteq(u)$ then $v \in \interp{F(u)}$.
Therefore for each $u,v\in N$, $\interp{F(u)} =
\simset_\subseteq(u)$. So defining $\form:=F$ we obtain the desired
result.
\end{proof}
\else \fi

\iffullversion
\begin{thm}
Algorithm~\ref{alg:schematic-GRE} with input $\gG = \tup{N,\to,l}$
terminates in time
$O(\size{{\to}}^2\cdot\size{N}^3+\size{N}\cdot\size{\rg\ l})$.
\end{thm}
\fi

\iffullversion
\begin{proof}
For a naive implementation, the main loop executes at most
$\size{N}^2$ may times and to find $u,v,w$ in
line~\ref{alg:line:loop-begin} we need
$O(\size{N}\cdot\size{{\to}}^2)$ many steps. The running time of the
loop body is absorbed by this last quantity. Hence, in total, the
execution time of the main loop is
$O(\size{{\to}}^2\cdot\size{N}^3)$. For each $v\in N$, we need
$O(\size{\rg\ l})$ many steps for lines~\ref{alg:line:init1}
and~\ref{alg:line:init2}. So in total, the initialization takes
$O(\size{N}\cdot\size{\rg\ l})$ many steps.
\end{proof}
\else \fi

\iffullversion
\begin{algorithm} \small
\io

\ForEach{$v\in \Delta$}{ $\prevS(v):=\Delta$

\If{for all binary $r$, $\su{r}{v}=\emptyset$}{$S(v):=\{u\in \Delta
\mid \unary(v)\ \subseteq \unary(u) \}$

$F(v):=\bigwedge\unary(v)$ } \Else{ let $r$ be a binary relation
such that $\su{r}{v}\not=\emptyset$

 $S(v):=\{u\in \Delta \colon
\unary(v) \subseteq \unary(u)\wedge\su{r}{u}\not=\emptyset \}$

$F(v):=\diam\top\wedge\bigwedge P(v)$ }}

\While{$\exists\ v\in \Delta:S(v)\not=\prevS(v)$}{

$\{\ I_1:\mbox{\bf assert\ }(\forall v)\ S(v)\subseteq\prevS(v)\ \}$

$\{\ I_2:\mbox{\bf assert\ }(\forall r,u,v,w)\
[(u,v)\in\interp{r}\wedge w\in
S(u)]\Rightarrow[\su{r}{w}\cap\prevS(v)\not=\emptyset]\ \}$

\ForEach{\mbox{\rm binary relation $r$ and $u\in \pr{r}{v}$}}{

$remove := \pr{r}{\prevS(v)}\setminus\pr{r}{S(v)}$

$S(u):=S(u)\setminus remove$

\If{$\diam F(v)$ is not a conjunct of $F(u)$}{ $F(u):=F(u)\wedge
\diam F(v)$\label{alg:line:loop-body-end-2} } }

$\prevS(v)=S(v)$

} \caption{\small refined $\EL$-similarity and
\posre}\label{alg:refined-GRE}
\end{algorithm}
\fi

Algorithm \ref{alg:schematic-GRE} can be easily modified to
calculate the \ELAN-RE of each simulator set $\simset_{\ELAN}$ by
adjusting the initialization: replace $\subseteq$ by $=$ in the
initialization of $S(v)$ and initialize $F(v)$ as
$\bigwedge\left(\unary(v)\cup\overline \unary(v)\right)$, where
$\overline \unary(v)=\{\lnot p\mid v\notin\interp{p}\}$.

With a naive implementation Algorithm \ref{alg:schematic-GRE}
executes in time $O(\size{\Delta}^3\times$
$\size{\interp{\cdot}}^2)$
%\fixme{Santi: chequear la complejidad}
providing a polynomial solution to the \EL and \ELAN-GRE problems.
\cite{HHK95} show a more involved version of
Algorithm~\ref{alg:schematic-gen-sim} with lower complexity, which
can be adapted in a similar way to compute $F(v)$. We shall skip the
details.

%Given $\gG = \tup{N,\to,l}$,
%the
%
%$F$, the set of \posre, and $S$, the simulator sets s.t.\ $(\forall
%v\in \Delta)\ \interp{F(v)}=S(v)=\simset_\EL(v)$

%
\begin{thm}\label{thm:complexity-EL-GRE}
The \EL/\ELAN-GRE problems over $\gM=\tup{\Delta,\interp{\cdot}}$
have complexity $O(\size{\Delta}\times\size{\interp{\cdot}})$.
\end{thm}

Theorem~\ref{thm:complexity-EL-GRE} answers a question left open by
\cite{AKS08}: the \EL-GRE problem can be solved in polynomial time.
Note that the above result assumes a convenient representation of
formulas as directed acyclic graphs (for $O(1)$ formula
construction). In section \ref{sec:size} we will take a look at this
in more detail.

%Theorem~\ref{thm:complexity-EL-GRE} shows that there are efficient
%GRE algorihtms for \EL and \ELAN; languages which are --as discussed
%in \S~\ref{sec:choosinglanguage}-- generally simple to realize. The
%methods presented here, together with previous results
%of~\cite{AKS08} indicate that algorithms for computing
%$\+L$-simulator sets can, in many cases, be adapted to solve the
%$\+L$-GRE problem at no extra complexity cost.

We have not addressed the issue of preferences with respect to the
use of certain relations, and moreover, we have presented our
algorithms as close as possible to the original proposal
of~\cite{HHK95} which makes them prioritize unary relations over
binary relations. The latter can be avoided by representing unary
relations as binary relations (cf.~\S\ref{sec:krahmer}).
%as is done by \cite{Krahmer2003}.
Certain control on preferences can then be introduced by taking them
in consideration instead of the non-deterministic choice of
differentiating elements made in the main loop of the algorithm. But
despite these modifications, the algorithms based on simulator sets
seem to offer less room for implementing preferences than the ones
we will discuss in \S~\ref{sec:krahmer}.

%But first we will briefly discuss a downside of the presented algorithm.

\newcommand{\instFun}[2]{\texttt{#1}\ensuremath{_{#2}}\xspace}

\section{GRE via building simulated models}\label{sec:krahmer}

We now revisit the algorithm presented by \cite{Krahmer2003},
identify its underlying notion of expressiveness, and extend it to
accommodate other notions. For reasons of space we assume the reader
is familiar with this algorithm and refer her to that article for
further information.

We must first note that scenes are encoded in that article in a
slightly different way: there, graphs have only labels on edges, and
non-relational attributes such as \emph{type} or \emph{color} are
represented by loops (e.g., $\aSmall(a,a)$). While our presentation
is, arguably, conceptually cleaner, it forces us to treat the atomic
and relational cases separately.

The second thing to note is that the output of their algorithm is a
\emph{connected subgraph} $H$ of the input graph $G$ that includes
the target $u$ among its nodes. This means that the algorithm does
not fit in the definition of $\+L$-GRE we presented in
\S\ref{sec:technical}.

Now, $H$ must be such that every \emph{subgraph isomorphism}%
\footnote{%
 A \emph{subgraph isomophism} between $G_1$ and $G_2$ is an isomorphism between $G_1$
 and a subgraph of $G_2$.
} $f $ between $H$ and $G$ satisfies $f(u) = u$. On relational
models, subgraph isomorphism corresponds to \EPFOL-simulations,
which makes explicit the notion of expressiveness that was used.
Indeed, from the output $H$ of this algorithm, one can easily build
a \EPFOL-formula that univoquely describes the target $u$, as is
shown in Algorithm~\ref{alg:build-form-epfol}. Observe that if
$\FOL$-simulations were used instead, we would have to include also
which unary and binary relations \emph{do not hold} in $H$.

\begin{algorithm}\small
\SetKwFunction{buildForm}{buildFormula$_\EPFOL$} \caption{\small
$\buildForm(H,v)$\label{alg:build-form-epfol}} \tcp{let $H =
\tup{\cset{a_1\ldots a_n}, \interp{\cdot}}$, $v=a_1$} $\gamma$ :=
$\displaystyle \bigwedge_{\mathclap{a_i \neq a_j}} (x_i \not\approx
x_j) \land \bigwedge_{\mathclap{(a_i,a_j) \in \interp{r}}}
r(x_i,x_j) \land \bigwedge_{\mathclap{a_i \in \interp{p}}}p(x_i) $\;
\Return{$\exists x_2\ldots \exists x_n . \gamma$}\;
\end{algorithm}

Having made explicit the notion of sameness underlying the algorithm
of \cite{Krahmer2003} and, with it, the logical language associated
to it, we can proceed to generalize the algorithm, as shown in
Algorithm~\ref{alg:makeRE}. This algorithm is parametric on $\+L$;
to make it concrete, one needs to provide appropriate versions of
\instFun{buildFormula}{\+L} and \instFun{extend}{\+L}. In order to
make the discussion of the differences with the original algorithm
simpler, we list the code for \instFun{buildFormula}{\EPFOL} and
\instFun{extend}{\EPFOL} in Algorithms~\ref{alg:build-form-epfol}
and~\ref{alg:extend-epfol}.

\begin{algorithm}\small
\SetKwFunction{makeRE}{makeRE$_\+L$}
\SetKwFunction{findGraph}{find$_\+L$}
\SetKwFunction{init}{init_$\+L$}
\SetKwFunction{buildForm}{buildFormula$_\+L$}

\caption{\small \makeRE($v$)\label{alg:makeRE}} $v_H$ := \emph{new
node}\; $\tup{H,f}$ := $\tup{\tup{\cset{v_H},\emptyset,\emptyset},
\cset{v_H \mapsto v}}$\; $H'$ := \findGraph($v_H, \bot, H, f$)\;
\Return{\buildForm($H',v_H$)}
\end{algorithm}

\begin{algorithm} \small
\SetKwFunction{findGraph}{find$_\+L$} \SetKwFunction{cost}{cost}
\SetKwFunction{matchGraph}{match$_\+L$}
\SetKwFunction{extendGraph}{extend$_\+L$}

\caption{\small \findGraph($v_H, \mathit{best}, H,f$)}

\If{$\mathit{best} \neq \bot \land \cost(\mathit{best}) \leq
\cost(H)$}{\Return $\mathit{best}$} $\mathit{distractors}$ :=
$\cset{n \mid n \in \Delta_G \land n \neq v \land v_H \simul{\+L}
n}$\; \If{$\mathit{distractors} = \emptyset$}{\Return $H$}
\ForEach{$\tup{H',f'} \in \extendGraph(H,f)$}{
  $I$ := \findGraph($v_H,\mathit{best},H', f'$)\;
  \If{$\mathit{best} = \bot \lor \cost(I) \leq \cost(\mathit{best})$}{$\mathit{best} := I$}
} \Return{$\mathit{best}$}
\end{algorithm}

Notice that \instFun{makeRE}{\+L} computes not only a graph $H$ but
also an $\+L$-simulation $f$. In the case of \EPFOL, $H$ is a
subgraph of $G$ and, therefore, $f$ is the trivial identity function
$\mathit{id(x)} = x$. We will see the need for $f$ when discussing
the case of less expressive logics like \EL. Observe also that in
\instFun{extend}{\EPFOL} we follow the notation by
\cite{Krahmer2003} and write, for a model $\+M = \tup{\Delta,
\interp{\cdot}}$,  $\+M + p(u)$ to denote the model $\tup{\Delta
\cup \cset{u},\interp{\cdot}'}$ such that $\interp{p}' = \interp{p}
\cup \cset{u}$ and $\interp{q}' = \interp{q}$ when $q \neq p$.
Similarly, $\+M + r(u,v)$ denotes the model $\tup{\Delta \cup
\cset{u,v},\interp{\cdot}'}$ such that $\interp{r}' = \interp{r}
\cup \{(u,v)\}$ and $\interp{q}' = \interp{q}$ when $q \neq r$. It
is clear, then, that this function is returning all the
\emph{extensions} of $H$ by adding a missing attribute or relation
to $H$, just like is done in the original algorithm.

\begin{algorithm}\small
\SetKwFunction{extendGraph}{extend$_\EPFOL$}

\caption{\small $\extendGraph(H,f)$\label{alg:extend-epfol}}

 $a$ := $\cset{H {+} p(u) \mid u \in \Delta_H, u \in \interp{p}_G  \setminus \interp{p}_H}$\;
 $b$ := $\cset{H {+} r(u,v) \mid u \in \Delta_H, (u,v) \in \interp{r}_G \setminus \interp{r}_H}$\;
 \Return{$(a \cup b) \times \cset{\mathit{id}}$}
\end{algorithm}

We now discuss a version of this algorithm for \EL. The first thing
to note is that one could, in principle, just use
\instFun{extend}{\EPFOL} also for \EL. Indeed, since
\instFun{find}{\EL} uses an \EL-simulation to compute the set of
\emph{distractors} (in the terminology of \cite{Krahmer2003}), the
output of this function would be a subgraph $H$ of $G$ such that for
every \EL-simulation $\sim$, $u \sim v$ iff $u = v$. The problem is
this subgraph may contain cycles and, as was observed in
\S\ref{sec:technical}, they cannot be tell apart using \EL. The
upshot is that we might be unable to realize the outcome of such
function.

A well-known result establishes that every relational model $\+M$ is
equivalent, with respect to \EL-formulas\footnote{Actually, the
result holds even for \ALC formulas.}, to the \emph{unraveling} of
$\+M$ (cf.~\cite{BRV01}).  That is, any model and its unraveling
satisfy exactly the same \EL formulas.  Morever, the unraveling of
$\+M$ is always a tree, and as we show in
Algorithm~\ref{alg:build-form-el}, it is straightforward to extract
a suitable \EL-formula from a tree.

\begin{algorithm} \small
\SetKwFunction{buildForm}{buildFormula$_\EL$} \caption{\small
$\buildForm(H,v)$\label{alg:build-form-el}} \mbox{{\bf requires} $H$
to be a tree}

$\gamma$ := $\cset{\exists r.\buildForm(H,u) \mid (v,u) \in
\interp{r}} $\; \Return{$ (\bigwedge\gamma) \land (\bigwedge_{v \in
\interp{p}}p)$}\;
\end{algorithm}

Therefore, we need \instFun{extend}{\EL} to return all the possible
extensions of $H$ by either adding a new proposition or a new edge
that is present in the unraveling of $G$ but not in $H$. This is
shown in Algorithm~\ref{alg:extend-el}.

Observe that the behavior of \instFun{find}{\EL} is quite sensible
to the cost function $f$ employed. For instance, on cyclic models,
an $f$ that does not guarantee the unraveling is explored in a
breadth-first way may lead to non-termination (since
\instFun{find}{\EL} may loop exploring an infinite branch).

It is also possible to use modal model-theoretical results to put a
bounds check that avoids generating an unraveling of infinite depth
when there is no possible referring expression, but we will not go
into the details for reasons of space.

\begin{algorithm}\small
\SetKwFunction{extendGraph}{extend$_\EL$} \caption{\small
$\extendGraph(H,f)$\label{alg:extend-el}}

$a$ :=  $\cset{\tup{H {+} p(u),f} \mid u \in \Delta_H, u \in
\interp{p}_G {-} \interp{p}_H}$\; $b$ := $\emptyset$\; \ForEach {$u
\in \Delta_G$}{
  \ForEach{$u_H \in \Delta_H / (f(u_h),u) \in \interp{r}_G$}{
    \If{$\forall v . ((u_H,v) \in \interp{r}_H \Rightarrow f(v) \neq u)$}{
      $n$ := \emph{new node}\;
      $b$ := $b \cup \cset{\tup{H + r(u_H,n),f[n \mapsto u]}}$\;
    }
  }
  }
  \Return{$a \cup b$}
\end{algorithm}

As a final note on complexity, although the set of \EL-distractors
may be computed more efficiently than \EPFOL-distractors, we cannot
conclude that \instFun{find}{\EL} is more efficient than
\instFun{find}{\EPFOL} in general: the model built in the first case
may be exponentially larger (it is an unraveling, after all).

\section{Combining GRE methods}\label{sec:combining}

An appealing feature of formulating the GRE problem modulo
expressivity is that one can devise general strategies that combine
$\+L$-GRE algorithms. We illustrate this with an example.

The algorithms based on $\+L$-simulator sets like the ones in
\S\ref{sec:simulation} simultaneously compute referring expressions
for every object in the domain, and do this for many logics in
polynomial time. This is an interesting property when one
anticipates the need of referring to a large number of elements.
However, this family of algorithms is not as flexible in terms of
implementing preferences as those  in \S\ref{sec:krahmer}.

There is a simple way to obtain an algorithm that is a compromise
between these two techniques. Let $A_1$ and $A_2$ be two procedures
that solve the $\+L$-GRE problem based on the techniques of
\S\ref{sec:simulation} and \S\ref{sec:krahmer}, respectively.  One
can first compute an $\+L$-RE for every possible object using $A_1$
and then (lazily) replace the calculated RE for $u$ with $A_2(u)$
whenever the former does not conform to some predefined criterion.
But one can do better.

Since $A_1$ computes, for a given $\+M =
\tup{\Delta,\interp{\cdot}}$, the set $\simset(u)$ for every $u \in
\Delta$, one can  build in polynomial time, using the output of
$A_1$, the model $\+M_{\+L} = \tup{\cset{[u] \mid u \in \Delta},
\interp{\cdot}_{\+L}}$, such that:
$$
\begin{array}{l@{\;=\;}l}
\setlength{\abovedisplayskip}{0pt}%
\setlength{\abovedisplayshortskip}{0pt}%
[u] & \cset{v \mid u \simul{\+L} v \text{ and } v \simul{\+L} u}\\
%\interp{p}_{\+L} &= \cset{[u] \mid u \in \interp{p}}\\
\interp{r}_{\+L} & \cset{([u_1]\ldots [u_n]) \mid (u_1\ldots u_n)
\in \interp{r}}
\end{array}
$$
$\+M_{\+L}$ is known as \emph{the $\+L$-minimization of $\+M$}. By a
straightforward induction on $\gamma$ one can verify that
$(u_1\ldots u_n) \in \interp{\gamma}$ iff $([u_1]\dots [u_n]) \in
\interp{\gamma}_{\+L}$ and this implies that $\gamma$ is a $\+L$-RE
for $u$ in $\+M$ iff it is a $\+L$-RE for $[u]$ in $\+M_{\+L}$.

If $\+M$ has a large number of indistinguishable elements (using
$\+L$), then
 $\+M_{\+L}$ will be much smaller than $\+M$. Since the computational complexity of
 $A_2$ depends on the size of $\+M$, for very large scenes, one should compute
 $A_2([u])$ instead.

%
% \subsection{Incremental expressivity}
%
%It may appear from what we have discussed so far that one has to pick an expressivity $\+L$
%in advance and stick to it. That is not necessarily true.  Let $\+L_0$ be a sublanguage or $\+L_1$
%(e.g., \EL and \EPFOL, respectively) and assume one prefers $\+L_0$-REs, although for some
%elements of the domain $\+L_0$ may not be enough.

%In order to obtain a RE for an element $u$ in $\+M$, one can first try an $\+L_0$-GRE algorithm and
%obtain a formula $\gamma_0$. If $\interp{\gamma_0} = \cset{u}$ then we are done. If not, instead of
%finding a $\+L_1$-RE for $u$ in $\+M$, one can run a $\+L_1$-GRE algorithm for $u$ in
%$\+M_{\gamma_0}$, where $\+M_{\gamma_0}$\ldots\fixme{hace falta un algoritmo que calcule deltas}

\section{On the size of referring expressions} \label{sec:size}

The expressive power of a language $\+L$ determines if there is an
$\+L$-RE for an element $u$. But observe that when $u$ can be
described in $\+L$, it may also influence the \emph{size} of the
\emph{shortest} $\+L$-RE. Intuitively, with more expressive power we
are able to `see' more differences and therefore have more resources
at hand to build a shorter formula.

A natural question is, then, whether we can characterize the
relative size of the $\+L$-REs for a given $\+L$. That is, if we can
give (tight) upper bounds for the size of the shortest $\+L$-REs for
the elements of an arbitrary model $\+M$, as a function of the size
of $\+M$.

%\fixme{Incluir ejemplo de S/U y FO (Kams Result).}

For the case of one of the most expressive logics considered in this
article, \EPFOL, the answer follows from algorithm
\instFun{makeRE}{\EPFOL} in \S\ref{sec:krahmer}. Indeed, if an
\EPFOL-RE exists, it is computed by \instFun{buildFormula}{\EPFOL}
from a model $H$ that is not bigger than the input model. It is easy
to see that this formula is linear in the size of $H$ and, therefore
the size of any \EPFOL-RE is $O(\size{\Delta} +
\size{\interp{\cdot}})$. It is not hard to see that this upper bound
holds for \FOL-REs too (cf.~\S\ref{sec:krahmer} for details).

Although \instFun{buildFormula}{\EL} also returns a formula that is
linear in the size of the tree-model $H$, $H$ could be, in
principle, exponentially larger than the input model. We can use
this to give an exponential upper bound for the size of the shortest
\EL-RE, but is it tight?

One is tempted to conclude from Theorem~\ref{thm:complexity-EL-GRE}
that the size of shortest \EL-RE is $O(\size{\Delta} \times
\size{\interp{\cdot}})$, but there is a pitfall.
Theorem~\ref{thm:complexity-EL-GRE} assumes that formulas are
represented as a DAG and guarantees this DAG is polynomial in the
size of the input model. One can easily reconstruct
 (the syntax tree of) the formula from the DAG, but this, in principle, may lead
 to a exponential blow-up (the result will be a exponentially larger formula,
 but composed of only a polynomial number of different subformulas).

As the following example shows, it is indeed possible to obtain an
\EL-formula that is exponentially larger when expanding the DAG
representation generated by Algorithm~\ref{alg:schematic-GRE}.

\begin{ex}\label{ex:bad-length}
Consider a language with only one binary relation $r$, and let
$\gM=\tup{\Delta,\interp{\cdot}}$ where $\Delta=\{1,2,\dots n\}$ and
$(i,j)\in\interp{r}$ iff $i<j$. Algorithm~\ref{alg:schematic-GRE}
initializes $F(j)=\top$ for all $j\in \Delta$. Suppose the following
choices in the execution: For $i=1,\dots n-1$, iterate $n-i$ times
picking $v=w=n-i+1$ and successively $u=n-i,\dots 1$. It can be
shown that each time a
formula $F(j)$ %($1\leq j<n$)
 is updated, it changes from $\phi$ to
$\phi\wedge\diam\phi$ and hence it doubles its size. Since $F(1)$ is
updated $n-1$ many times, the size of $F(1)$ is greater than $2^n$.

\iffullversion Suppose that in the first $n-1$ iterations we
successively choose $v=w=n$ and $u=n-1,n-2,\dots,1$. That is, we
discover that $n$ does not simulate $n-1,n-2,\dots,1$. We end up
with $F(1)=\dots=F(n-1)=\top\wedge\diam\top$ and
$S(1)=\dots=S(n-1)=W\setminus\{n\}$. Suppose that in the following
$n-2$ iterations we successively choose $v=w=n-1$ and
$u=n-2,n-3,\dots,1$, namely, we discover that $n-1$ does not
simulate $n-2,\dots,1$. We end up with
$F(i)=(\top\wedge\diam\top)\wedge \diam(\top\wedge\diam\top)$ and
$S(i)=\Delta\setminus\{n,n-1\}$, for $1 \le i \le n-1$. Following
this choice scheme, each time $F(1)$ is updated, it changes from
$\varphi$ to $\varphi\wedge\diam\varphi$. Since $F(1)$ is updated
$n-1$ many times, the size of $F(1)$ is
$O(2^n)=O(2^{\size{\Delta}})$. \fi
\end{ex}

The large $\+L$-RE of Example~\ref{ex:bad-length} is due to an
unfortunate (non-deterministic) choice of elements.
Example~\ref{ex:not-that-bad-length} shows that another execution
 leads to a quadratic RE (but notice the
shortest one is linear).

\begin{ex}\label{ex:not-that-bad-length}
%Let $\gM$ be the model of Example~\ref{ex:bad-length}.
Suppose now that in the first $n-1$ iterations we successively
choose $v=w=n-i$ and $u=v-1$ for $i=0\dots n-2$. It can be seen that
for convenient choices, $F(1)$ is of size
$O(n^2)$.%
%
%
%We end up with $F(n)=\top$ and $F(i)=\top\wedge\diam F(i+1)$ for
%$i=1,\dots,n-1$. Up to this point, the size of $F(1)$ is linear in
%$n$. Of course, the algorithm does not stop here, but the reader can
%verify that if we successively choose $u=1$ and $v=w=n,n-1,\dots,2$
%we end up with $F(1)$ of size $O(n^2)$.
\end{ex}

We are yet unable to answer whether the exponential bound for the
size of the minimum \EL-RE is tight. We conjecture no polynomial
bound can be given, though. In any case, it seems clear that not
only existence of RE but relative lengths should be taken into
account when considering the trade-off between expressive powers.

%Note that executions showed in examples
%
%note that in both examples, element $1$ can be described with the
%formula $(\exists r)^{(n-1)}.\top$, which is of size $O(n)$.

\iffullversion
\begin{ex}
Let $\gM$ be the model of example~\ref{ex:bad-length}. Suppose that
in the first $n-1$ iterations we successively choose $v=w=n,u=n-1$;
$v=w=n-1,u=n-2$; $\dots$; $v=w=2,u=1$. That is, we discover that
$n-i+1$ does not simulate $n-i$, for successive $i=1,2,\dots,n-1$.
We end up with $F(n)=\top$ and $F(i)=\top\wedge\diam F(i+1)$ for
$i=1,\dots,n-1$. Up to this point, the size of $F(1)$ is linear in
$n$. But this $F(1)$ is not the final value because we still have to
discover that $3,\dots,n$ do not simulate $1$. The reader may verify
that if we successively choose $u=1$ and $v=w=n,n-1,\dots,2$ we end
up with $F(1)$ of size quadratic in $n$.
\end{ex}
\fi

\iffullversion \fixme{This example is a bit too specific.} Let $\gG$
be any finite graph. If $u,v$ of $\gG$ are bisimilar then for all
formula $\varphi\in\pos$, $u \in \interp{\varphi}$ iff $v \in
\interp{\varphi}$. Observe that in this case, $F(u)$ and $F(v)$
computed by Algorithm~\ref{alg:schematic-GRE} need not necessarily
be equal.

\begin{ex}
PONER EJEMPLO DE ESTO ULTIMO.
\end{ex}
\fi

\section{Conclusions}\label{conclusions}

There is some notion of expressiveness underlying the formulation of
every GRE problem. This ``expressiveness'' can be formally measured
in terms of a logical language or, dually, a simulation relation
between models. In this article we have discussed making the notion
of expressiveness involved an explicit parameter of the GRE problem,
unlike usual practice.

We have taken an abstract view, defining the ``$\+L$-GRE problem'';
and though we considered various possible choices for $\+L$, we did
not argue for any of them. Instead, we tried to make explicit the
trade-off involved in the selection of a particular $\+L$. This, we
believe, depends heavily on the given context.

By making expressiveness explicit, we can transfer general knowledge
and results from the well-developed field of \emph{computational
logics}. This was exemplified in \S\ref{sec:simulation} and
\S\ref{sec:krahmer} where we were able to turn known GRE algorithms
into \emph{families} of algorithms that may deal analogously with
different logical languages. We also applied this in
\S\ref{sec:combining} to devise new heuristics.

Arguably, an explicit notion of expressiveness also provides a
cleaner interface, either between the content-determination and
surface realization modules or between two collaborating
content-determination modules. An instance of the latter was
exhibited in \S\ref{sec:combining}.

As a future line of research, one may want to avoid sticking to a
fixed $\+L$ but instead favor an incremental approach in which
features of a more expressive language $\+L_1$ are used only when
$\+L_0$ is not enough to distinguish certain element.

%\fixme{Podemos generalizar el algoritmo para conjuntos? Mencionar algoritmo de Piazza}

%To finish this section, observe that the algorithms introduced here
%can also be used to compute referring expressions for \emph{sets} of
%elements. Let $\+L$ be \EL or \ELAN. For any $v\in\Delta$ we define
%the $\+L$-class of $v$ as
%%\fixme{$[v]_{\+L}$ pide demasiado. alcanza con que sean todos los $u$ tq $v \simul{\+L} u$, no?}
%$$
%[v]_{\+L}=\{u\in\Delta\mid u\in\simset_{\+L}(v)\wedge
%v\in\simset_{\+L}(u)\}.
%$$
%A set $T\subseteq\Delta$ has an $\+L$-RE iff $T=[u]_{\+L}$ for some
%$u\in\Delta$. In case $T=[u]_{\+L}$ for some $u$ then for any
%$v\in[u]_{\+L}$, $F(v)$ is a $\+L$-RE for $T$.
%
%Since computing the $\+L$-classes of $\Delta$ is polynomial in
%$\size{\Delta}$, Theorem \ref{thm:complexity-EL-GRE} implies the
%following:

%\fixme{ARREGLAR} To finish this section, observe that the algorithms
%introduced here can also be used to compute referring expressions
%for some \emph{sets} of elements. Let $\+L$ be \EL or \ELAN. A set
%$T\subseteq\Delta$ has an $\+L$-RE iff $T=\simset_{\+L}(u)$ for some
%$u\in\Delta$. In case $T=\simset_{\+L}(u)$ for some $u$ then $F(u)$
%is a $\+L$-RE for $T$. In fact, $F(v)$ also is, for any $v$ such
%that $u\in\simset_{\+L}(v)\wedge v\in\simset_{\+L}(u)$.

%Hence by Theorem \ref{thm:complexity-EL-GRE} we have:

%\begin{cor}
%The problem of generating the \EL and \ELAN referring expressions of
%sets of elements given a finite model
%%$\gM=\tup{\Delta,\interp{\cdot}}$
%can be solved in polynomial time.
%\end{cor}

\bibliographystyle{acl}
\bibliography{plan}

\end{document}